\documentclass[aps,pre,twocolumn,superscriptaddress]{revtex4-2}
\usepackage{amssymb}
\usepackage{amsmath,bm}
\usepackage{graphicx,color}
\usepackage{bbm}
\usepackage[bottom]{footmisc}
\usepackage{multirow}
\usepackage{xcolor}
\usepackage{xpatch} 
\makeatletter   
\usepackage{xpatch} 
 \usepackage{booktabs}

\newcommand{\B}[1]{{\bm{#1}}}

\begin{document}
\title{Symmetry Breaking and Energy Dissipation in the Mechanical Response of Amorphous Solids}
\author{Itamar Procaccia} 
\affiliation{Sino-Europe Complex Science Center, School of Mathematics, North University of China, Shanxi, Taiyuan 030051, China.}
\affiliation{Hangzhou International Innovation Institute, Beihang University, Hangzhou 311115, China}
\affiliation{Dept. of Chemical Physics, The Weizmann Institute of Science, Rehovot 76100, Israel}
\author{Tuhin Samanta}
\affiliation{Dept. of Chemical Physics, The Weizmann Institute of Science, Rehovot 76100, Israel}

\begin{abstract}
Dissipation, or the loss of energy conservation, is necessarily related to symmetry breaking. Here we study the symmetry breaking that is responsible for dissipation in the mechanical response of amorphous solids to quasi-static strains. To this aim we consider carefully just one cycle of strain, to reveal the source of dissipation. In general the response can be conservative or dissipative, with a sharp transition between these options as a function of preparation parameters, accompanied by translation and rotational symmetry breaking and the onset of screening. We choose an example for which the mesoscopic theory can be solved exactly, and the microscopic physics can be revealed by numerical simulations. The mechanism of energy dissipation (when it exists) is explored in detail, showing that in this case it is due to the breaking of rotational symmetry.
\end{abstract}
\maketitle

{\bf Introduction}: Engineers were concerned with the
fatigue that results from cyclic applied strains for many decades \cite{54Mil,66Dow,89DBS,94KC}. Physicists and material scientists use cyclic driving to study mechanical properties of materials, like self-organization \cite{08CCGP,13RLR}, memory effects \cite{14PK,14FFS,18BHPRZ,19MSDR,23SL}, emergent computational capabilities \cite{21BH,24LTKVJBH, 23KH}, and yielding \cite{20DVS,22BHP}. In amorphous solids, every cycle can result in dissipation, rearrangement of configurations and a complicated exploration of the complex energy landscape that is associated with microscopic disorder. Nevertheless, notwithstanding the immense effort and considerable progress, the precise source of dissipation mechanisms that play a role in cyclic strain protocols, including the power-law convergence of the amount of dissipation with the number of cycles \cite{18BHPRZ,26SBPHL}, still remain elusive. From the Noether's theorem we know that a loss of conservation is associated with a continuous symmetry breaking; the question is which symmetry breaking is at work in the dissipative response to cyclically driven amorphous solids. Especially in quasi-static protocols where time is not involved, the question is particularly relevant.

\begin{figure}
	\includegraphics[width=0.45\textwidth]{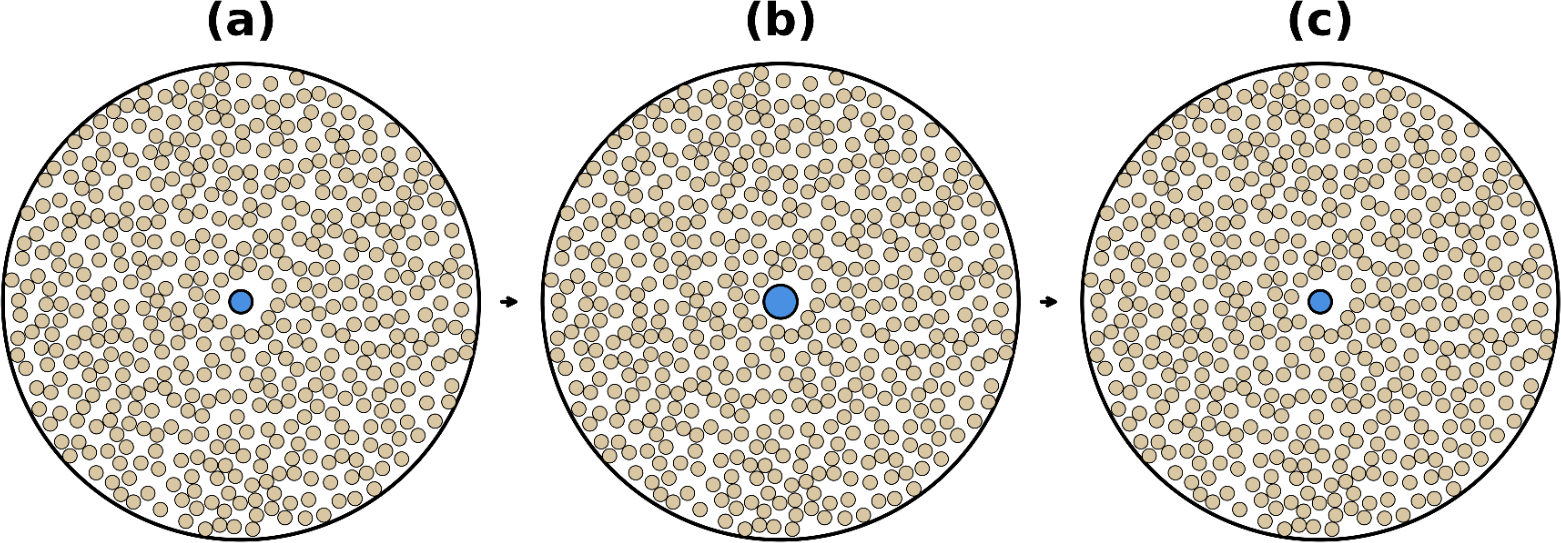}
	\caption{Schematic diagram of the cycle. (a) Equilibrated before inflation. (b) Equilibrated after inflation. (c) Equilibrated after deflation.}
	\label{scheme}  
\end{figure}

In this Letter we propose that light can be shed on the physics of cyclic strain by considering just one cycle, and studying what is happening from both microscopic and mesoscopic viewpoints. We choose a cycle for which the mesoscopic theory can be described analytically, whereas its microscopic aspects are revealed by numerical simulations. The system that we have in mind is by now well studied \cite{21LMMPRS,22KMPS,24JPS,24FHKKP,25IPTSPNAS} - an amorphous solid that is contained in an annulus, between an outer boundary of radius $r_{\rm out}$ and inner boundary of radius $r_{\rm in}$. First the material is brought to mechanical equilibrium, and then the inner boundary is inflated by some chosen percentage and the system is brought again to mechanical equilibrium. A new aspect of the present Letter is that after equilibration the inner boundary is deflated to its original radius, and the system is again brought to mechanical equilibrium. The entire process is one cycle, which is shown graphically in Fig.~\ref{scheme}.  In particular we want to shed light on when and how dissipation sets in, and when it does, which broken symmetry is responsible for the loss of energy conservation during a cycle. 
In spite of the elementary nature of this protocol, we will see that it involves rich and complex physics, including screening, symmetry breaking, dissipation and phase transitions. Due to its relative simplicity it can serve as an analog of the harmonic oscillator in molecular physics - an analytically tractable example with lots to learn from.

{\bf The cycle}:  The system that we employ below consists of  assemblies of small frictionless disks confined in a circular two-dimensional area with a fixed outer circular wall, prepared in mechanical equilibrium, with a desired packing fraction $\Phi$ above the jamming packing fraction $\Phi_j$.  Open-source code (LAMMPS) \cite{95Pli} is used to perform the simulations. The studied cycle begins with a configuration of $N = 80,000$ bi-dispersed disks of mass $m = 1$, randomly placed in a circular area with a radius $r_{\text{out}}$. Half of the small disks have a radius $R_1 = 0.45$ and the other half have a radius $R_2 = 0.65$. One large disk of size $R_2$ is not placed randomly, but is rather fixed in the center of the coordinates. The normal contact force is Hertzian with force constant $k_n = 2 \times 10^5$, following the Discrete Element Method of Ref.~\cite{79CS}. 

The system is relaxed to mechanical equilibrium by solving Newton’s second law of motion with damping. This process is carried out until the resultant force on each disk is minimized to values smaller than $10^{-6}$. The energy and pressure $E_b$ and  $P_b$ (with the subscript standing for ``before" inflation) are then calculated.

Once a mechanically stable configuration is reached, the central disk is inflated $r_{\rm in}\to r_{\rm in}+d_0$ by some percentage \(5\% \ldots 25\% \), and subsequently, mechanical equilibration follows, and the resulting displacement field $\B d(r,\theta)$ is measured. Afterward, the central particle is deflated back to its original radius, the mechanical equilibration procedure is performed again, and the system's energy and pressure, termed as $E_f$ and $P_f$ (final) is calculated. In this step, we further measure the resulting displacement field between the equilibrated configuration (after inflation) and the final equilibrated configuration  (after deflation) . Finally, the energy difference between the final (after deflation) and the initial (before inflation) configurations $\Delta E\equiv E_b-E_f$ is calculated and reported here. The cycle is schematically represented in Fig.~\ref{scheme}. One of our aims is to understand when and why the cycle is dissipative with $\Delta E\ne 0$. 

{\bf Simulation Results:} In spite of the simplicity of this cycle, the observed results can vary widely, primarily due to the sensitive dependence on the pressure. There exists a critical pressure $P_c$, such that for $P_b>P_c$ the response to the inflation is quasi-elastic, and the whole cycle is basically conservative, without dissipation. At $P_c$ one has a transition in the mechanical response of the system, associated with divergent angular correlation length and non-trivial exponents, and see  Refs. \cite{24JPS,25FYDP} for a physical explanation of the existence of a critical pressure.  It had been shown before \cite{21LMMPRS,22BMP,22KMPS,22MMPRSZ,23CMP,24KP} that in this regime the displacement field after inflation satisfies the classical equation $\B \nabla \cdot \B \sigma=0$, which when translated to from the stress $\B \sigma$ to the displacement $\B d$ reads
\begin{equation}
 \Delta \mathbf{d} + \left(\tilde\lambda + 1 \right) \nabla \left(\nabla\cdot \mathbf{d}\right) = 0 \ , 
	\label{d1}
\end{equation}
where $\tilde \lambda$ is the ratio of the Lam\`e coefficients, $\tilde \lambda \equiv \lambda/\mu$.
Plastic events in this regime, if they occur, are small in extent and can at most renormalize the elastic moduli $\lambda$ and $\mu$ \cite{21LMMPRS,23CMP}. Consequently, in the present radial geometry one predicts that the radial component of the displacement field, $d_r\equiv \B d(r,\theta)\cdot \B {\hat r}$, $\B {\hat r}\equiv \B r/r$, is 
\begin{eqnarray}
	d_r (r) = d_0 \frac{r_{\text{in}} \left(r^2-r_{\text{out}}^2\right)}{r \left(r_{\text{in}}^2-r_{\text{out}}^2\right)} \ .
	\label{renelas}
\end{eqnarray} 

\begin{figure}[h!]
	\includegraphics[width=0.35\textwidth]{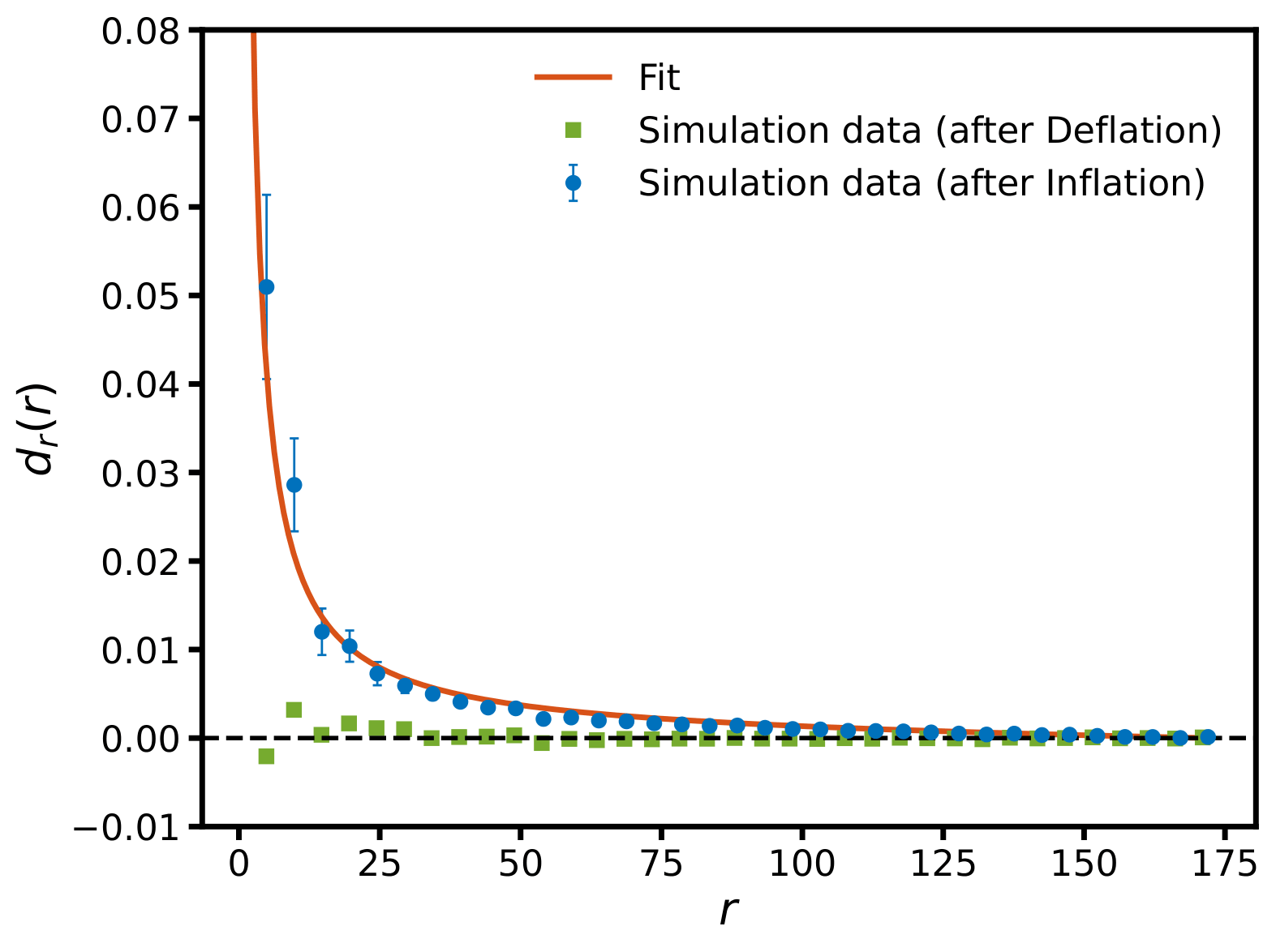}
	\caption{The comparisons of the angle-averaged radial displacement field, \( d_r(r) \) with its corresponding analytical solution Eq.~(\ref{renelas}) (depicted as solid lines). Here $P=30$, and the inflation is $25\%$}
		\label{normal} 
		\end{figure}

This prediction is borne out by the numerical simulations as can be seen for example in Fig.~\ref{normal}.
Here the pressure is $P=30$ and the inflation is 25\%, i.e. $d_0 =1.25 R_2$. The angle-averaged radial component of the displacement field is in excellent agreement with Eq.~(\ref{renelas}), the displacement field (computed with respect to the initial equilibrated configuration) after deflation is almost zero, and the energy returns to its original value, $\Delta E/E_b\approx O(10^{-5})$. The same conclusion pertains to all the high pressure ($P>P_c$) simulations, for all the inflation values that we examined between 5-25\%.

A world of differences opens up when the pressure is lower than $P_c$. Here plastic avalanches create topological charges, both quadrupolar and dipolar, these result in translational and rotational symmetry breaking that lead to dissipation. The displacement field satisfies now a modified equation, 
\begin{equation}\label{L0}
	\bf  \Delta  {d} + (\tilde{\lambda} +1)\nabla (\nabla \cdot  {d}) +\Gamma  {d} =0,
\end{equation}
where $\Gamma$ has the following form \cite{23CSWDM,24FHKKP,25KPS},
\begin{align}\label{L1}
	\bf \Gamma=
	\begin{bmatrix}
		\bf    \kappa_{e}^{2} & \bf\mp\kappa_{o}^{2}  \\
		\bf   \pm \kappa_{o}^{2} &\bf \kappa_{e}^{2} 
	\end{bmatrix}.
\end{align}
The appearance of the odd screening parameter $\kappa_{o}$ is crucial for the rest of this Letter, as explained next. Its physical meaning is a rotation of the local dipole field with respect to the local displacement field. With $\kappa_{o}=0$ the local dipole field $\B P(\B r)$ is necessarily co-linear with the local displacement field $\B d(\B r)$, but oppositely signed, i.e. $\B P(\B r)=-\kappa_e^2 \B d$.  With the existence of $\kappa_{o}$ the local dipole field is rotated by an angle $\theta$ with respect to the displacement field. This angle is continuous, does not have a fixed value, representing rotational symmetry breaking. It can also be positive or negative, depending on the sight of $\kappa_{o}^{2}$ in Eq.~\ref{L1}. But in a given realization it is either positive or negative, and one can refer to this as Chiral symmetry breaking. But the continuous nature of $\theta$ is providing the connection to energy dissipation as is shown below. In Ref.~\cite{25FYDP} the consequences of rotational symmetry breaking are studied and we recommend the interested reader to examine that reference.

The screening parameters, both the diagonal (which lead to translational symmetry breaking) and off-diagonal (which lead to rotational symmetry breaking), are a-priori predictable to high precision from the properties of the solutions of the governing equation (\ref{L0}), and see Refs. \cite{24JPS,25FYDP} for details. 

In studying Eq.~(\ref{L0}) the tangential component of the displacement field is as interesting as the radial component, and we denote
\begin{equation}\label{L2}
	{\B d} = d_{r}(r,\theta)\hat{r} + d_{\theta}(r,\theta)\hat{\theta}.
\end{equation}

To demonstrate the fundamental change in mechanical response, we present here simulations of our system at pressure $P=0.41$, for four different sizes of inflation, i.e. 5\%, 10\%, 15\% and 25\%.
\begin{figure}[h!]
	\includegraphics[width=0.30\textwidth]{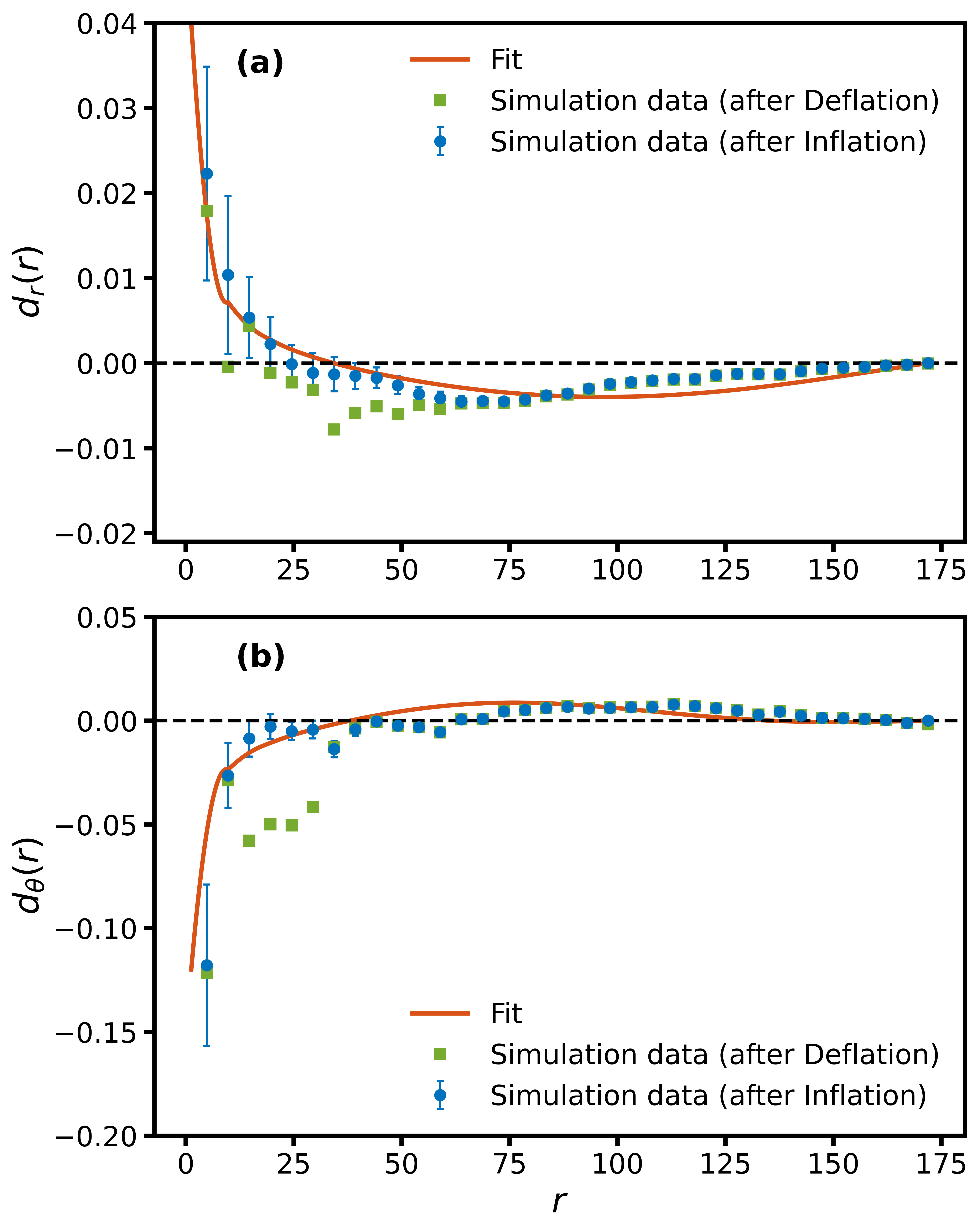}
	\includegraphics[width=0.29\textwidth]{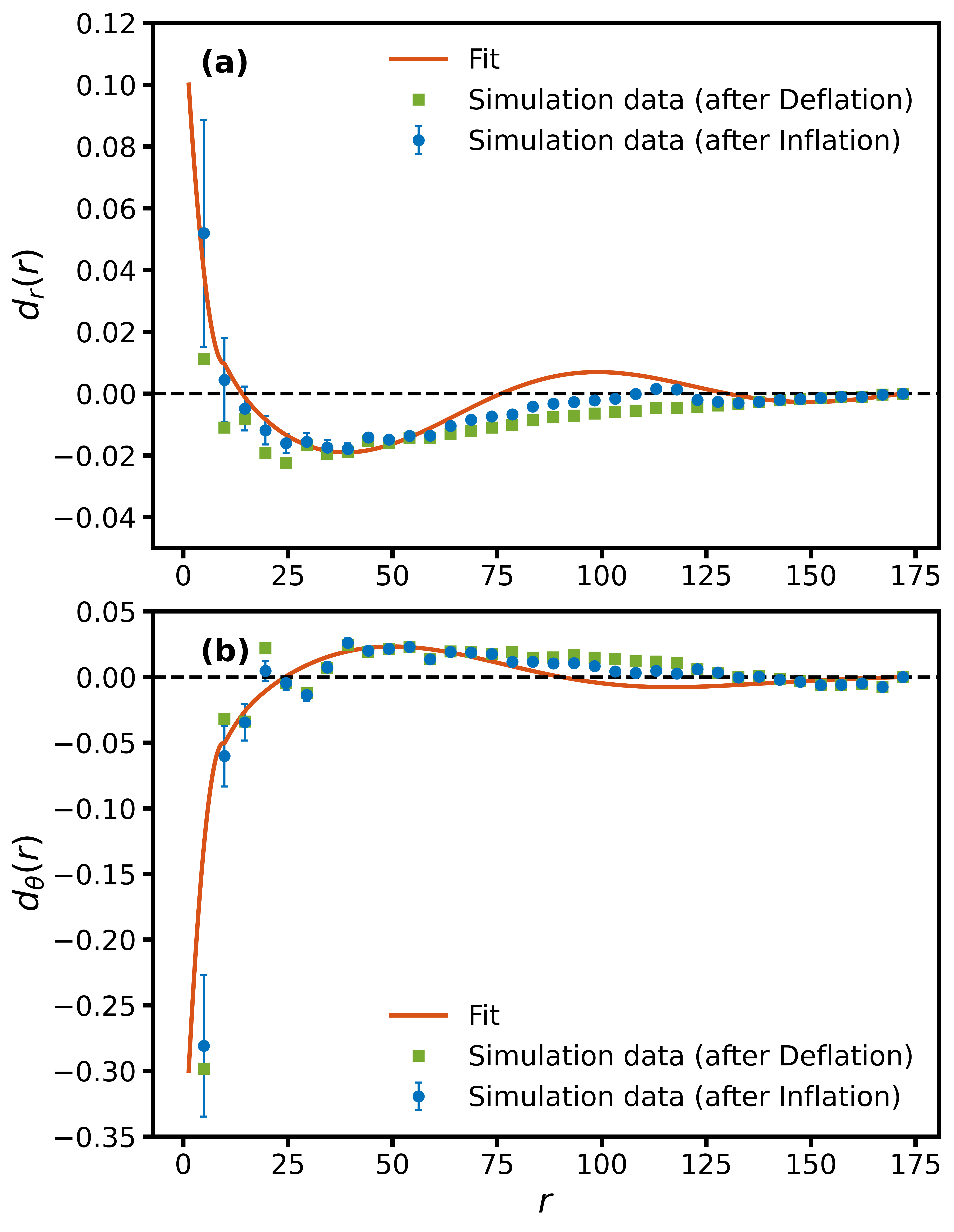}
	\caption{Radial and tangential angle averaged displacement fields as a function of $r$. Here $P=0.41$. Upper two panels, $d_0=1.05R_2$. Lower two panels, $d_0=1.10R_2$.}
	\label{anom1}
\end{figure}
\begin{figure}
	\includegraphics[width=0.30\textwidth]{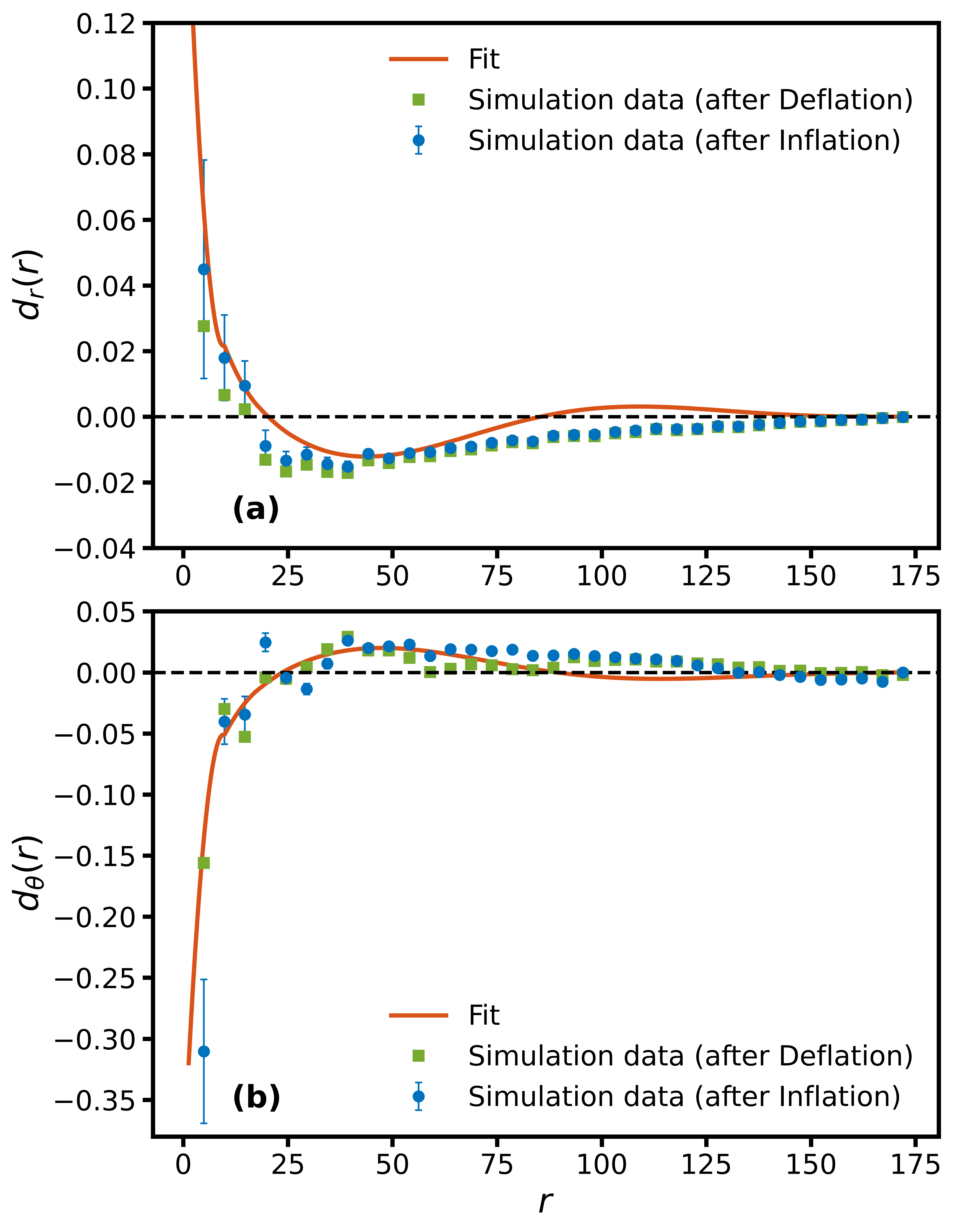}
	\includegraphics[width=0.29\textwidth]{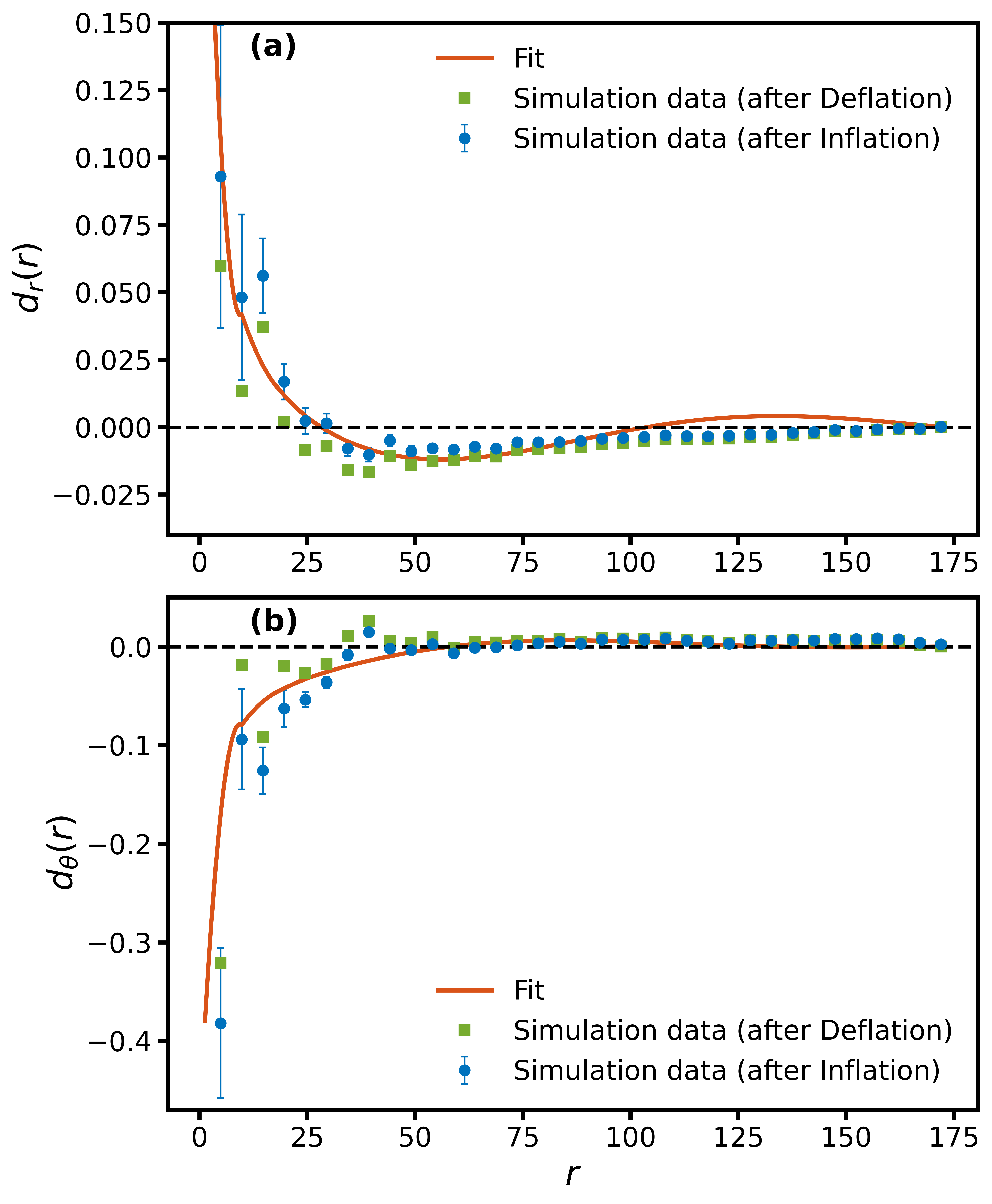}
	\caption{Radial and tangential angle averaged displacement fields as a function of $r$. Here $P=0.41$. Upper two panels, $d_0=1.15R_2$. Lower two panels, $d_0=1.25R_2$.}
	\label{anom2}
\end{figure}
The results are shown in Figs.~\ref{anom1} and \ref{anom2}. In every pair of panels (a) and (b) we show, for a given value of the inflation, the angle-averaged radial and tangential components of the displacement field, with the red continuous line being ``best" fit among the solutions of Eq.~(\ref{L0}), The blue dots are the simulation results, and the green squares are the displacement field after deflation, this time computed with respect to the equilibrated configuration after inflation. The screening parameters $\kappa_e$ and $\kappa_o$ can be predicted theoretically to very good precision \cite{25KPS}, and see Table~\ref{table} for the quantitative comparison. 
\begin{table}
	\resizebox{\linewidth}{!}{%
		\begin{tabular}{cccccccc}
			\toprule
			\textbf{Inflation (\%)} & \boldmath$E_f$ & \boldmath$P_f$ & \boldmath$\kappa_e$ & \boldmath$\kappa_o$ & \boldmath$\kappa^{\rm Th}_e$ & \boldmath$\kappa^{\rm Th}_o$  & \boldmath$\Delta E~\%$ \\
			\midrule
			\text{5.0}  & 27.82 & 0.38  & 0.053 & 0.086 & 0.049 & 0.080 & 7.94 \\
			\text{10.0} & 25.02 & 0.346 & 0.074 & 0.140 & 0.078 & 0.135 & 17.21  \\
			\text{15.0} & 25.24 & 0.344 & 0.071 & 0.140 & 0.067 & 0.127 & 16.48 \\
			\text{25.0} & 23.70 & 0.322 & 0.056 & 0.098 & 0.057 & 0.092 & 21.58 \\
			\bottomrule
		\end{tabular}%
	}
	\caption{The final energy and pressure, the fitted and predicted screening parameters and the percentage dissipation for four different values of inflation.}
	\label{table}
\end{table}
In this table we present, in addition to the final energy and pressure, the fitted and predicted values of the screening parameters (the latter with subscript ``Th"), and the percentage change in energy. How the screening parameters are predicted theoretically is explained in Refs.~\cite{24JPS,25FYDP,25KPS}.

{\bf Discussion:} the main lessons to learn are (i) when the pressure is below $P_c$ one expects and finds anomalous mechanical response that is faithfully described by the solutions of Eq.~(\ref{L0}). (ii) The anomalous response, which is triggered by plastic avalanches, is accompanied with energy dissipation. (iii) 
The most important clue that we learn from the numerical results shown in Figs.~\ref{anom1} and \ref{anom2} is that the displacement field after inflation relative to the initial equilibrated configuration (green data points) is very closely the same as the displacement field after deflation with respect to the inflated equilibrated configuration (blue points). This will help us to identify the symmetry breaking that is responsible for the dissipation.

To see how this works we need to recall that upon inflation we load the system since we increase the pressure. Then $\Delta E_1>0$ the work done can be estimated as the integral of the force times the displacement. Since the force, $\B \nabla \cdot \B \sigma$ is in the present case $-\mu\Gamma^{\alpha\beta} d_\alpha$, the work done is the integral 
\begin{equation}
	\Delta E_1 =-\frac{\mu}{A}\int_\Omega d^2 r \Gamma^{\alpha\beta} d_\alpha(\B r) d_\beta (\B r) \ , 
\end{equation}
	where $A$ is the area between $r_{in}$ and $r_{out}$.
Upon deflation, and {\em because we find that the displacement fields are the same, characterized by the same screening parameters}, the work released is 
\begin{eqnarray}
	\Delta E_2 &=&\frac{\mu}{A}\int_\Omega d^2 r \Gamma^{\alpha\beta} d_\alpha (\B r)d_\beta (\B r)\nonumber\\& =&\frac{\mu}{A}\int_\Omega d^2 r \Gamma^{\beta\alpha} d_\beta(\B r) d_\alpha (\B r)\ ,
\end{eqnarray}
where we exchanged the order of the dummy indices $\alpha$ and $\beta$. Finally,
the total dissipation is then $\Delta E\equiv \Delta E_1+\Delta E_2$,
\begin{equation}
\Delta E = \frac{\mu}{A}\int_\Omega d^2 r  [\Gamma^{\beta\alpha}-\Gamma^{\alpha\beta}] d_\alpha (\B r) d_\beta	(\B r)= \kappa_o^2 I\ .
\label{dissvskap}
\end{equation}
Here we denoted by $I$ the integral once $\kappa_0^2$ is extracted. 

\begin{figure}
	\includegraphics[width=0.45\textwidth]{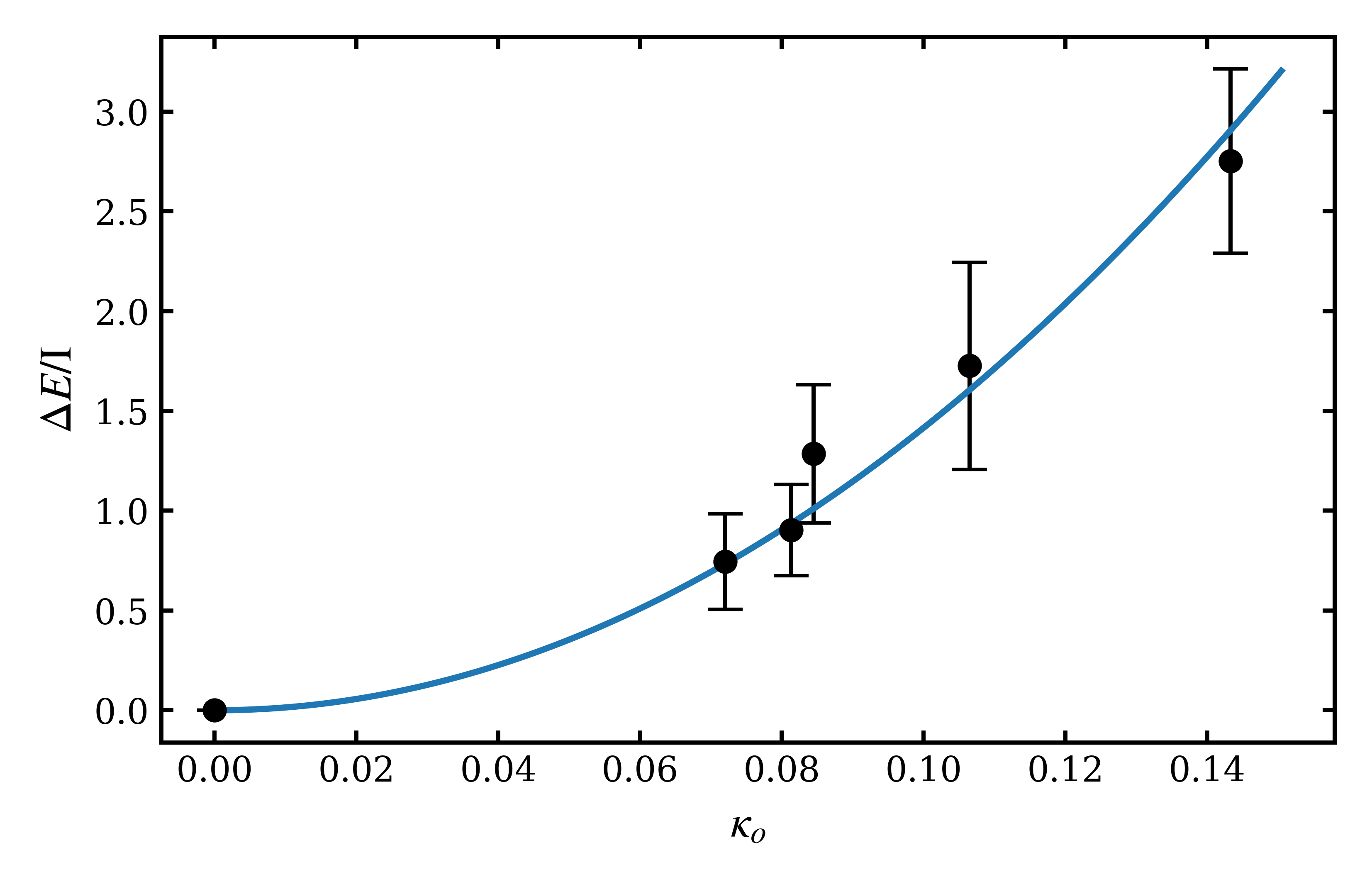}
	\caption{The energy dissipation (normalized by the integral $I$, cf. Eq.~(\ref{dissvskap}), as a function of $\kappa_o$. Note that the theory expects zero dissipaton when $\kappa_o=0$, so the point (0,0) is included in the data. The prediction of Eq.~(\ref{dissvskap}) which is represented with the continuous line, is supported by the data.}
	\label{propkap}
	\end{figure}

The obvious conclusion is that dissipation in the present case is due to the rotational symmetry breaking rather than the translation symmetry breaking. This interesting conclusion can be directly assessed in our numerical simulations, cf. Fig.~\ref{propkap}. To quantify the dissipative response, we consider five distinct geometries characterized by
\begin{eqnarray}
(r_{\rm in}, r_{\rm out}) = &&(0.65, 172.00),(0.665,176.51), (0.675, 179.51)\nonumber \\  &&(0.70, 187.30),(0.75, 202.51).
\end{eqnarray}
For each geometry, the pressure is $P = 0.41$, and we follow the same inflation--deflation protocol with a deformation amplitude $d_0 = 1.15\,R_2$. For each geometry, four independent realizations are considered in order to obtain statistically averaged quantities.
For each realization, the normalized energy $\Delta E/I$ is calculated. We also determine the corresponding odd-screening parameter $\kappa_o$ for each case. The average values $\langle \Delta E/I \rangle$ and $\langle \kappa_o \rangle$ are calculated across realizations for each geometric case.
Fig.~\ref{propkap} represents $\langle \Delta E/I \rangle$ as a function of the  odd screening exponent $\kappa_o$.

In summary, for the present example of perhaps the simplest possible cyclic strain, the mechanical response may or may not
include dissipation, depending on the material condition. Above a critical pressure, the cycle ends without dissipation, the response is quasi-elastic, and no symmetry is broken. Below the critical pressure, plastic response leads to the creation of quadrupolar and dipolar effective charges in the displacement field, which is always accompanied by the breaking of symmetries, both translation
and rotational. The breaking of symmetry leads to the loss of energy conservation. The numerical simulations provide evidence that points to the rotational symmetry breaking, which is associated with the odd part of the screening tensor $\Gamma$
(cf. Eq.~(\ref{L1})), as the fundamental reason for the loss of energy conservation.

\section*{Acknowledgments}

This work was supported by the joint grant between the Israel Science Foundation and the National Natural Science Foundation of China, and by the Minerva Foundation, Munich, Germany. T.S. acknowledges financial support from WSoS and the use of computational resources provided by CHEMFARM.

\section*{Author Information}

\noindent
Correspondence should be addressed to: \\
Itamar Procaccia (itamar.procaccia@weizmann.ac.il) \\
Tuhin Samanta (tuhin.samanta@weizmann.ac.il)
\bibliography{ALL.anomalous}

\begin{thebibliography}{31}%
\makeatletter
\providecommand \@ifxundefined [1]{%
 \@ifx{#1\undefined}
}%
\providecommand \@ifnum [1]{%
 \ifnum #1\expandafter \@firstoftwo
 \else \expandafter \@secondoftwo
 \fi
}%
\providecommand \@ifx [1]{%
 \ifx #1\expandafter \@firstoftwo
 \else \expandafter \@secondoftwo
 \fi
}%
\providecommand \natexlab [1]{#1}%
\providecommand \enquote  [1]{``#1''}%
\providecommand \bibnamefont  [1]{#1}%
\providecommand \bibfnamefont [1]{#1}%
\providecommand \citenamefont [1]{#1}%
\providecommand \href@noop [0]{\@secondoftwo}%
\providecommand \href [0]{\begingroup \@sanitize@url \@href}%
\providecommand \@href[1]{\@@startlink{#1}\@@href}%
\providecommand \@@href[1]{\endgroup#1\@@endlink}%
\providecommand \@sanitize@url [0]{\catcode `\\12\catcode `\$12\catcode
  `\&12\catcode `\#12\catcode `\^12\catcode `\_12\catcode `\%12\relax}%
\providecommand \@@startlink[1]{}%
\providecommand \@@endlink[0]{}%
\providecommand \url  [0]{\begingroup\@sanitize@url \@url }%
\providecommand \@url [1]{\endgroup\@href {#1}{\urlprefix }}%
\providecommand \urlprefix  [0]{URL }%
\providecommand \Eprint [0]{\href }%
\providecommand \doibase [0]{https://doi.org/}%
\providecommand \selectlanguage [0]{\@gobble}%
\providecommand \bibinfo  [0]{\@secondoftwo}%
\providecommand \bibfield  [0]{\@secondoftwo}%
\providecommand \translation [1]{[#1]}%
\providecommand \BibitemOpen [0]{}%
\providecommand \bibitemStop [0]{}%
\providecommand \bibitemNoStop [0]{.\EOS\space}%
\providecommand \EOS [0]{\spacefactor3000\relax}%
\providecommand \BibitemShut  [1]{\csname bibitem#1\endcsname}%
\let\auto@bib@innerbib\@empty
\bibitem [{\citenamefont {Miles}(1954)}]{54Mil}%
  \BibitemOpen
  \bibfield  {author} {\bibinfo {author} {\bibfnamefont {J.~W.}\ \bibnamefont
  {Miles}},\ }\bibfield  {title} {\bibinfo {title} {On structural fatigue under
  random loading},\ }\href@noop {} {\bibfield  {journal} {\bibinfo  {journal}
  {Journal of the Aeronautical Sciences}\ }\textbf {\bibinfo {volume} {21}},\
  \bibinfo {pages} {753} (\bibinfo {year} {1954})}\BibitemShut {NoStop}%
\bibitem [{\citenamefont {Dowell}(1966)}]{66Dow}%
  \BibitemOpen
  \bibfield  {author} {\bibinfo {author} {\bibfnamefont {E.~H.}\ \bibnamefont
  {Dowell}},\ }\bibfield  {title} {\bibinfo {title} {Nonlinear oscillations of
  a fluttering plate.},\ }\href {https://doi.org/10.2514/3.3658} {\bibfield
  {journal} {\bibinfo  {journal} {AIAA Journal}\ }\textbf {\bibinfo {volume}
  {4}},\ \bibinfo {pages} {1267} (\bibinfo {year} {1966})}\BibitemShut
  {NoStop}%
\bibitem [{\citenamefont {Dietmann}\ \emph {et~al.}(1989)\citenamefont
  {Dietmann}, \citenamefont {Bhongbhibhat},\ and\ \citenamefont
  {Schmid}}]{89DBS}%
  \BibitemOpen
  \bibfield  {author} {\bibinfo {author} {\bibfnamefont {H.}~\bibnamefont
  {Dietmann}}, \bibinfo {author} {\bibfnamefont {T.}~\bibnamefont
  {Bhongbhibhat}},\ and\ \bibinfo {author} {\bibfnamefont {A.}~\bibnamefont
  {Schmid}},\ }\bibfield  {title} {\bibinfo {title} {Multiaxial fatigue
  behaviour of steels under in-phase and out-of-phase loading including
  different wave forms and frequencies},\ }in\ \href@noop {} {\emph {\bibinfo
  {booktitle} {ICBMFF3}}}\ (\bibinfo {year} {1989})\BibitemShut {NoStop}%
\bibitem [{\citenamefont {Klevtsov}\ and\ \citenamefont {Crane}(1994)}]{94KC}%
  \BibitemOpen
  \bibfield  {author} {\bibinfo {author} {\bibfnamefont {I.}~\bibnamefont
  {Klevtsov}}\ and\ \bibinfo {author} {\bibfnamefont {R.}~\bibnamefont
  {Crane}},\ }\bibfield  {title} {\bibinfo {title} {{Random Thermal Stress
  Oscillations and Fatigue Life Estimation for Steam Generator Tubes}},\ }\href
  {https://doi.org/10.1115/1.2929563} {\bibfield  {journal} {\bibinfo
  {journal} {Journal of Pressure Vessel Technology}\ }\textbf {\bibinfo
  {volume} {116}},\ \bibinfo {pages} {110} (\bibinfo {year}
  {1994})}\BibitemShut {NoStop}%
\bibitem [{\citenamefont {Corte}\ \emph {et~al.}(2008)\citenamefont {Corte},
  \citenamefont {Chaikin}, \citenamefont {Gollub},\ and\ \citenamefont
  {Pine}}]{08CCGP}%
  \BibitemOpen
  \bibfield  {author} {\bibinfo {author} {\bibfnamefont {L.}~\bibnamefont
  {Corte}}, \bibinfo {author} {\bibfnamefont {P.~M.}\ \bibnamefont {Chaikin}},
  \bibinfo {author} {\bibfnamefont {J.~P.}\ \bibnamefont {Gollub}},\ and\
  \bibinfo {author} {\bibfnamefont {D.~J.}\ \bibnamefont {Pine}},\ }\bibfield
  {title} {\bibinfo {title} {Random organization in periodically driven
  systems},\ }\href@noop {} {\bibfield  {journal} {\bibinfo  {journal} {Nature
  Physics}\ }\textbf {\bibinfo {volume} {4}},\ \bibinfo {pages} {420} (\bibinfo
  {year} {2008})}\BibitemShut {NoStop}%
\bibitem [{\citenamefont {Regev}\ \emph {et~al.}(2013)\citenamefont {Regev},
  \citenamefont {Lookman},\ and\ \citenamefont {Reichhardt}}]{13RLR}%
  \BibitemOpen
  \bibfield  {author} {\bibinfo {author} {\bibfnamefont {I.}~\bibnamefont
  {Regev}}, \bibinfo {author} {\bibfnamefont {T.}~\bibnamefont {Lookman}},\
  and\ \bibinfo {author} {\bibfnamefont {C.}~\bibnamefont {Reichhardt}},\
  }\bibfield  {title} {\bibinfo {title} {Onset of irreversibility and chaos in
  amorphous solids under periodic shear},\ }\href@noop {} {\bibfield  {journal}
  {\bibinfo  {journal} {Physical Review E}\ }\textbf {\bibinfo {volume} {88}},\
  \bibinfo {pages} {062401} (\bibinfo {year} {2013})}\BibitemShut {NoStop}%
\bibitem [{\citenamefont {Paulsen}\ and\ \citenamefont {Keim}(2024)}]{14PK}%
  \BibitemOpen
  \bibfield  {author} {\bibinfo {author} {\bibfnamefont {J.~D.}\ \bibnamefont
  {Paulsen}}\ and\ \bibinfo {author} {\bibfnamefont {N.~C.}\ \bibnamefont
  {Keim}},\ }\bibfield  {title} {\bibinfo {title} {Mechanical memories in
  solids, from disorder to design},\ }\href@noop {} {\bibfield  {journal}
  {\bibinfo  {journal} {Annual Review of Condensed Matter Physics}\ }\textbf
  {\bibinfo {volume} {16}} (\bibinfo {year} {2024})}\BibitemShut {NoStop}%
\bibitem [{\citenamefont {Fiocco}\ \emph {et~al.}(2014)\citenamefont {Fiocco},
  \citenamefont {Foffi},\ and\ \citenamefont {Sastry}}]{14FFS}%
  \BibitemOpen
  \bibfield  {author} {\bibinfo {author} {\bibfnamefont {D.}~\bibnamefont
  {Fiocco}}, \bibinfo {author} {\bibfnamefont {G.}~\bibnamefont {Foffi}},\ and\
  \bibinfo {author} {\bibfnamefont {S.}~\bibnamefont {Sastry}},\ }\bibfield
  {title} {\bibinfo {title} {Encoding of memory in sheared amorphous solids},\
  }\href@noop {} {\bibfield  {journal} {\bibinfo  {journal} {Physical review
  letters}\ }\textbf {\bibinfo {volume} {112}},\ \bibinfo {pages} {025702}
  (\bibinfo {year} {2014})}\BibitemShut {NoStop}%
\bibitem [{\citenamefont {Bandi}\ \emph {et~al.}(2018)\citenamefont {Bandi},
  \citenamefont {Hentschel}, \citenamefont {Procaccia}, \citenamefont {Roy},\
  and\ \citenamefont {Zylberg}}]{18BHPRZ}%
  \BibitemOpen
  \bibfield  {author} {\bibinfo {author} {\bibfnamefont {M.~M.}\ \bibnamefont
  {Bandi}}, \bibinfo {author} {\bibfnamefont {H.~G.~E.}\ \bibnamefont
  {Hentschel}}, \bibinfo {author} {\bibfnamefont {I.}~\bibnamefont
  {Procaccia}}, \bibinfo {author} {\bibfnamefont {S.}~\bibnamefont {Roy}},\
  and\ \bibinfo {author} {\bibfnamefont {J.}~\bibnamefont {Zylberg}},\
  }\bibfield  {title} {\bibinfo {title} {Training, memory and universal scaling
  in amorphous frictional granular matter},\ }\href
  {http://stacks.iop.org/0295-5075/122/i=3/a=38003} {\bibfield  {journal}
  {\bibinfo  {journal} {Europhys. Lett.}\ }\textbf {\bibinfo {volume} {122}},\
  \bibinfo {pages} {38003} (\bibinfo {year} {2018})}\BibitemShut {NoStop}%
\bibitem [{\citenamefont {Mungan}\ \emph {et~al.}(2019)\citenamefont {Mungan},
  \citenamefont {Sastry}, \citenamefont {Dahmen},\ and\ \citenamefont
  {Regev}}]{19MSDR}%
  \BibitemOpen
  \bibfield  {author} {\bibinfo {author} {\bibfnamefont {M.}~\bibnamefont
  {Mungan}}, \bibinfo {author} {\bibfnamefont {S.}~\bibnamefont {Sastry}},
  \bibinfo {author} {\bibfnamefont {K.}~\bibnamefont {Dahmen}},\ and\ \bibinfo
  {author} {\bibfnamefont {I.}~\bibnamefont {Regev}},\ }\bibfield  {title}
  {\bibinfo {title} {Networks and hierarchies: How amorphous materials learn to
  remember},\ }\href@noop {} {\bibfield  {journal} {\bibinfo  {journal}
  {Physical review letters}\ }\textbf {\bibinfo {volume} {123}},\ \bibinfo
  {pages} {178002} (\bibinfo {year} {2019})}\BibitemShut {NoStop}%
\bibitem [{\citenamefont {Shohat}\ and\ \citenamefont {Lahini}(2023)}]{23SL}%
  \BibitemOpen
  \bibfield  {author} {\bibinfo {author} {\bibfnamefont {D.}~\bibnamefont
  {Shohat}}\ and\ \bibinfo {author} {\bibfnamefont {Y.}~\bibnamefont
  {Lahini}},\ }\bibfield  {title} {\bibinfo {title} {Dissipation indicates
  memory formation in driven disordered systems},\ }\href@noop {} {\bibfield
  {journal} {\bibinfo  {journal} {Physical Review Letters}\ }\textbf {\bibinfo
  {volume} {130}},\ \bibinfo {pages} {048202} (\bibinfo {year}
  {2023})}\BibitemShut {NoStop}%
\bibitem [{\citenamefont {Bense}\ and\ \citenamefont {van Hecke}(2021)}]{21BH}%
  \BibitemOpen
  \bibfield  {author} {\bibinfo {author} {\bibfnamefont {H.}~\bibnamefont
  {Bense}}\ and\ \bibinfo {author} {\bibfnamefont {M.}~\bibnamefont {van
  Hecke}},\ }\bibfield  {title} {\bibinfo {title} {Complex pathways and memory
  in compressed corrugated sheets},\ }\href@noop {} {\bibfield  {journal}
  {\bibinfo  {journal} {Proceedings of the National Academy of Sciences}\
  }\textbf {\bibinfo {volume} {118}},\ \bibinfo {pages} {e2111436118} (\bibinfo
  {year} {2021})}\BibitemShut {NoStop}%
\bibitem [{\citenamefont {Liu}\ \emph {et~al.}(2024)\citenamefont {Liu},
  \citenamefont {Teunisse}, \citenamefont {Korovin}, \citenamefont {Vermaire},
  \citenamefont {Jin}, \citenamefont {Bense},\ and\ \citenamefont {van
  Hecke}}]{24LTKVJBH}%
  \BibitemOpen
  \bibfield  {author} {\bibinfo {author} {\bibfnamefont {J.}~\bibnamefont
  {Liu}}, \bibinfo {author} {\bibfnamefont {M.}~\bibnamefont {Teunisse}},
  \bibinfo {author} {\bibfnamefont {G.}~\bibnamefont {Korovin}}, \bibinfo
  {author} {\bibfnamefont {I.~R.}\ \bibnamefont {Vermaire}}, \bibinfo {author}
  {\bibfnamefont {L.}~\bibnamefont {Jin}}, \bibinfo {author} {\bibfnamefont
  {H.}~\bibnamefont {Bense}},\ and\ \bibinfo {author} {\bibfnamefont
  {M.}~\bibnamefont {van Hecke}},\ }\bibfield  {title} {\bibinfo {title}
  {Controlled pathways and sequential information processing in serially
  coupled mechanical hysterons},\ }\href@noop {} {\bibfield  {journal}
  {\bibinfo  {journal} {Proceedings of the National Academy of Sciences}\
  }\textbf {\bibinfo {volume} {121}},\ \bibinfo {pages} {e2308414121} (\bibinfo
  {year} {2024})}\BibitemShut {NoStop}%
\bibitem [{\citenamefont {Kwakernaak}\ and\ \citenamefont {van
  Hecke}(2023)}]{23KH}%
  \BibitemOpen
  \bibfield  {author} {\bibinfo {author} {\bibfnamefont {L.~J.}\ \bibnamefont
  {Kwakernaak}}\ and\ \bibinfo {author} {\bibfnamefont {M.}~\bibnamefont {van
  Hecke}},\ }\bibfield  {title} {\bibinfo {title} {Counting and sequential
  information processing in mechanical metamaterials},\ }\href@noop {}
  {\bibfield  {journal} {\bibinfo  {journal} {Physical Review Letters}\
  }\textbf {\bibinfo {volume} {130}},\ \bibinfo {pages} {268204} (\bibinfo
  {year} {2023})}\BibitemShut {NoStop}%
\bibitem [{\citenamefont {Das}\ \emph {et~al.}(2020)\citenamefont {Das},
  \citenamefont {Vinutha},\ and\ \citenamefont {Sastry}}]{20DVS}%
  \BibitemOpen
  \bibfield  {author} {\bibinfo {author} {\bibfnamefont {P.}~\bibnamefont
  {Das}}, \bibinfo {author} {\bibfnamefont {H.}~\bibnamefont {Vinutha}},\ and\
  \bibinfo {author} {\bibfnamefont {S.}~\bibnamefont {Sastry}},\ }\bibfield
  {title} {\bibinfo {title} {Unified phase diagram of reversible--irreversible,
  jamming, and yielding transitions in cyclically sheared soft-sphere
  packings},\ }\href@noop {} {\bibfield  {journal} {\bibinfo  {journal}
  {Proceedings of the National Academy of Sciences}\ }\textbf {\bibinfo
  {volume} {117}},\ \bibinfo {pages} {10203} (\bibinfo {year}
  {2020})}\BibitemShut {NoStop}%
\bibitem [{\citenamefont {Bhowmik}\ \emph
  {et~al.}(2022{\natexlab{a}})\citenamefont {Bhowmik}, \citenamefont
  {Hentchel},\ and\ \citenamefont {Procaccia}}]{22BHP}%
  \BibitemOpen
  \bibfield  {author} {\bibinfo {author} {\bibfnamefont {B.~P.}\ \bibnamefont
  {Bhowmik}}, \bibinfo {author} {\bibfnamefont {H.~G.~E.}\ \bibnamefont
  {Hentchel}},\ and\ \bibinfo {author} {\bibfnamefont {I.}~\bibnamefont
  {Procaccia}},\ }\bibfield  {title} {\bibinfo {title} {Fatigue and collapse of
  cyclically bent strip of amorphous solid},\ }\href
  {https://doi.org/10.1209/0295-5075/ac4ba5} {\bibfield  {journal} {\bibinfo
  {journal} {Europhysics Letters}\ }\textbf {\bibinfo {volume} {137}},\
  \bibinfo {pages} {46002} (\bibinfo {year} {2022}{\natexlab{a}})}\BibitemShut
  {NoStop}%
\bibitem [{\citenamefont {Shohat}\ \emph {et~al.}(2026)\citenamefont {Shohat},
  \citenamefont {Baconnier}, \citenamefont {Procaccia}, \citenamefont {van
  Hecke},\ and\ \citenamefont {Lahini}}]{26SBPHL}%
  \BibitemOpen
  \bibfield  {author} {\bibinfo {author} {\bibfnamefont {D.}~\bibnamefont
  {Shohat}}, \bibinfo {author} {\bibfnamefont {P.}~\bibnamefont {Baconnier}},
  \bibinfo {author} {\bibfnamefont {I.}~\bibnamefont {Procaccia}}, \bibinfo
  {author} {\bibfnamefont {M.}~\bibnamefont {van Hecke}},\ and\ \bibinfo
  {author} {\bibfnamefont {Y.}~\bibnamefont {Lahini}},\ }\bibfield  {title}
  {\bibinfo {title} {Aging of amorphous materials under cyclic strain},\ }\href
  {https://doi.org/10.1073/pnas.2515075123} {\bibfield  {journal} {\bibinfo
  {journal} {Proceedings of the National Academy of Sciences}\ }\textbf
  {\bibinfo {volume} {123}},\ \bibinfo {pages} {e2515075123} (\bibinfo {year}
  {2026})},\ \Eprint
  {https://arxiv.org/abs/https://www.pnas.org/doi/pdf/10.1073/pnas.2515075123}
  {https://www.pnas.org/doi/pdf/10.1073/pnas.2515075123} \BibitemShut {NoStop}%
\bibitem [{\citenamefont {Lema\^{\i}tre}\ \emph {et~al.}(2021)\citenamefont
  {Lema\^{\i}tre}, \citenamefont {Mondal}, \citenamefont {Moshe}, \citenamefont
  {Procaccia}, \citenamefont {Roy},\ and\ \citenamefont
  {Screiber-Re'em}}]{21LMMPRS}%
  \BibitemOpen
  \bibfield  {author} {\bibinfo {author} {\bibfnamefont {A.}~\bibnamefont
  {Lema\^{\i}tre}}, \bibinfo {author} {\bibfnamefont {C.}~\bibnamefont
  {Mondal}}, \bibinfo {author} {\bibfnamefont {M.}~\bibnamefont {Moshe}},
  \bibinfo {author} {\bibfnamefont {I.}~\bibnamefont {Procaccia}}, \bibinfo
  {author} {\bibfnamefont {S.}~\bibnamefont {Roy}},\ and\ \bibinfo {author}
  {\bibfnamefont {K.}~\bibnamefont {Screiber-Re'em}},\ }\bibfield  {title}
  {\bibinfo {title} {Anomalous elasticity and plastic screening in amorphous
  solids},\ }\href {https://doi.org/10.1103/PhysRevE.104.024904} {\bibfield
  {journal} {\bibinfo  {journal} {Phys. Rev. E}\ }\textbf {\bibinfo {volume}
  {104}},\ \bibinfo {pages} {024904} (\bibinfo {year} {2021})}\BibitemShut
  {NoStop}%
\bibitem [{\citenamefont {Kumar}\ \emph {et~al.}(2022)\citenamefont {Kumar},
  \citenamefont {Moshe}, \citenamefont {Procaccia},\ and\ \citenamefont
  {Singh}}]{22KMPS}%
  \BibitemOpen
  \bibfield  {author} {\bibinfo {author} {\bibfnamefont {A.}~\bibnamefont
  {Kumar}}, \bibinfo {author} {\bibfnamefont {M.}~\bibnamefont {Moshe}},
  \bibinfo {author} {\bibfnamefont {I.}~\bibnamefont {Procaccia}},\ and\
  \bibinfo {author} {\bibfnamefont {M.}~\bibnamefont {Singh}},\ }\bibfield
  {title} {\bibinfo {title} {Anomalous elasticity in classical glass formers},\
  }\href {https://doi.org/10.1103/PhysRevE.106.015001} {\bibfield  {journal}
  {\bibinfo  {journal} {Phys. Rev. E}\ }\textbf {\bibinfo {volume} {106}},\
  \bibinfo {pages} {015001} (\bibinfo {year} {2022})}\BibitemShut {NoStop}%
\bibitem [{\citenamefont {Jin}\ \emph {et~al.}(2024)\citenamefont {Jin},
  \citenamefont {Procaccia},\ and\ \citenamefont {Samanta}}]{24JPS}%
  \BibitemOpen
  \bibfield  {author} {\bibinfo {author} {\bibfnamefont {Y.}~\bibnamefont
  {Jin}}, \bibinfo {author} {\bibfnamefont {I.}~\bibnamefont {Procaccia}},\
  and\ \bibinfo {author} {\bibfnamefont {T.}~\bibnamefont {Samanta}},\
  }\bibfield  {title} {\bibinfo {title} {Intermediate phase between jammed and
  unjammed amorphous solids},\ }\href
  {https://doi.org/10.1103/PhysRevE.109.014902} {\bibfield  {journal} {\bibinfo
   {journal} {Phys. Rev. E}\ }\textbf {\bibinfo {volume} {109}},\ \bibinfo
  {pages} {014902} (\bibinfo {year} {2024})}\BibitemShut {NoStop}%
\bibitem [{\citenamefont {Fu}\ \emph {et~al.}(2024)\citenamefont {Fu},
  \citenamefont {Hentschel}, \citenamefont {Kaur}, \citenamefont {Kumar},\ and\
  \citenamefont {Procaccia}}]{24FHKKP}%
  \BibitemOpen
  \bibfield  {author} {\bibinfo {author} {\bibfnamefont {Y.}~\bibnamefont
  {Fu}}, \bibinfo {author} {\bibfnamefont {H.~G.~E.}\ \bibnamefont
  {Hentschel}}, \bibinfo {author} {\bibfnamefont {P.}~\bibnamefont {Kaur}},
  \bibinfo {author} {\bibfnamefont {A.}~\bibnamefont {Kumar}},\ and\ \bibinfo
  {author} {\bibfnamefont {I.}~\bibnamefont {Procaccia}},\ }\bibfield  {title}
  {\bibinfo {title} {Odd dipole screening in radial inflation},\ }\href
  {https://doi.org/10.1103/PhysRevE.110.065003} {\bibfield  {journal} {\bibinfo
   {journal} {Phys. Rev. E}\ }\textbf {\bibinfo {volume} {110}},\ \bibinfo
  {pages} {065003} (\bibinfo {year} {2024})}\BibitemShut {NoStop}%
\bibitem [{\citenamefont {Procaccia}\ and\ \citenamefont
  {Samanta}(2025)}]{25IPTSPNAS}%
  \BibitemOpen
  \bibfield  {author} {\bibinfo {author} {\bibfnamefont {I.}~\bibnamefont
  {Procaccia}}\ and\ \bibinfo {author} {\bibfnamefont {T.}~\bibnamefont
  {Samanta}},\ }\bibfield  {title} {\bibinfo {title} {Dipole-induced transition
  in 3 dimensions},\ }\href {https://doi.org/10.1073/pnas.2427273122}
  {\bibfield  {journal} {\bibinfo  {journal} {Proceedings of the National
  Academy of Sciences}\ }\textbf {\bibinfo {volume} {122}},\ \bibinfo {pages}
  {e2427273122} (\bibinfo {year} {2025})},\ \Eprint
  {https://arxiv.org/abs/https://www.pnas.org/doi/pdf/10.1073/pnas.2427273122}
  {https://www.pnas.org/doi/pdf/10.1073/pnas.2427273122} \BibitemShut {NoStop}%
\bibitem [{\citenamefont {Plimpton}(1995)}]{95Pli}%
  \BibitemOpen
  \bibfield  {author} {\bibinfo {author} {\bibfnamefont {S.}~\bibnamefont
  {Plimpton}},\ }\bibfield  {title} {\bibinfo {title} {{Fast Parallel
  Algorithms for Short-Range Molecular Dynamics}},\ }\href
  {https://doi.org/10.1006/jcph.1995.1039} {\bibfield  {journal} {\bibinfo
  {journal} {Journal of Computational Physics}\ }\textbf {\bibinfo {volume}
  {117}},\ \bibinfo {pages} {1} (\bibinfo {year} {1995})}\BibitemShut {NoStop}%
\bibitem [{\citenamefont {Cundall}\ and\ \citenamefont {Strack}(1979)}]{79CS}%
  \BibitemOpen
  \bibfield  {author} {\bibinfo {author} {\bibfnamefont {P.~A.}\ \bibnamefont
  {Cundall}}\ and\ \bibinfo {author} {\bibfnamefont {O.~D.~L.}\ \bibnamefont
  {Strack}},\ }\bibfield  {title} {\bibinfo {title} {A discrete numerical model
  for granular assemblies},\ }\href {https://doi.org/10.1680/geot.1979.29.1.47}
  {\bibfield  {journal} {\bibinfo  {journal} {Géotechnique}\ }\textbf
  {\bibinfo {volume} {29}},\ \bibinfo {pages} {47} (\bibinfo {year}
  {1979})}\BibitemShut {NoStop}%
\bibitem [{\citenamefont {Fu}\ \emph {et~al.}(2025)\citenamefont {Fu},
  \citenamefont {Jin}, \citenamefont {Pan},\ and\ \citenamefont
  {Procaccia}}]{25FYDP}%
  \BibitemOpen
  \bibfield  {author} {\bibinfo {author} {\bibfnamefont {Y.}~\bibnamefont
  {Fu}}, \bibinfo {author} {\bibfnamefont {Y.}~\bibnamefont {Jin}}, \bibinfo
  {author} {\bibfnamefont {D.}~\bibnamefont {Pan}},\ and\ \bibinfo {author}
  {\bibfnamefont {I.}~\bibnamefont {Procaccia}},\ }\bibfield  {title} {\bibinfo
  {title} {Long-range angular correlations of particle displacements at a
  plastic-to-elastic transition in jammed amorphous solids},\ }\href
  {https://doi.org/10.1103/PhysRevLett.134.178201} {\bibfield  {journal}
  {\bibinfo  {journal} {Phys. Rev. Lett.}\ }\textbf {\bibinfo {volume} {134}},\
  \bibinfo {pages} {178201} (\bibinfo {year} {2025})}\BibitemShut {NoStop}%
\bibitem [{\citenamefont {Bhowmik}\ \emph
  {et~al.}(2022{\natexlab{b}})\citenamefont {Bhowmik}, \citenamefont {Moshe},\
  and\ \citenamefont {Procaccia}}]{22BMP}%
  \BibitemOpen
  \bibfield  {author} {\bibinfo {author} {\bibfnamefont {B.~P.}\ \bibnamefont
  {Bhowmik}}, \bibinfo {author} {\bibfnamefont {M.}~\bibnamefont {Moshe}},\
  and\ \bibinfo {author} {\bibfnamefont {I.}~\bibnamefont {Procaccia}},\
  }\bibfield  {title} {\bibinfo {title} {Direct measurement of dipoles in
  anomalous elasticity of amorphous solids},\ }\href
  {https://doi.org/10.1103/PhysRevE.105.L043001} {\bibfield  {journal}
  {\bibinfo  {journal} {Phys. Rev. E}\ }\textbf {\bibinfo {volume} {105}},\
  \bibinfo {pages} {L043001} (\bibinfo {year}
  {2022}{\natexlab{b}})}\BibitemShut {NoStop}%
\bibitem [{\citenamefont {Mondal}\ \emph {et~al.}(2022)\citenamefont {Mondal},
  \citenamefont {Moshe}, \citenamefont {Procaccia}, \citenamefont {Roy},
  \citenamefont {Shang},\ and\ \citenamefont {Zhang}}]{22MMPRSZ}%
  \BibitemOpen
  \bibfield  {author} {\bibinfo {author} {\bibfnamefont {C.}~\bibnamefont
  {Mondal}}, \bibinfo {author} {\bibfnamefont {M.}~\bibnamefont {Moshe}},
  \bibinfo {author} {\bibfnamefont {I.}~\bibnamefont {Procaccia}}, \bibinfo
  {author} {\bibfnamefont {S.}~\bibnamefont {Roy}}, \bibinfo {author}
  {\bibfnamefont {J.}~\bibnamefont {Shang}},\ and\ \bibinfo {author}
  {\bibfnamefont {J.}~\bibnamefont {Zhang}},\ }\bibfield  {title} {\bibinfo
  {title} {Experimental and numerical verification of anomalous screening
  theory in granular matter},\ }\href
  {https://doi.org/https://doi.org/10.1016/j.chaos.2022.112609} {\bibfield
  {journal} {\bibinfo  {journal} {Chaos, Solitons and Fractals}\ }\textbf
  {\bibinfo {volume} {164}},\ \bibinfo {pages} {112609} (\bibinfo {year}
  {2022})}\BibitemShut {NoStop}%
\bibitem [{\citenamefont {Charan}\ \emph {et~al.}(2023)\citenamefont {Charan},
  \citenamefont {Moshe},\ and\ \citenamefont {Procaccia}}]{23CMP}%
  \BibitemOpen
  \bibfield  {author} {\bibinfo {author} {\bibfnamefont {H.}~\bibnamefont
  {Charan}}, \bibinfo {author} {\bibfnamefont {M.}~\bibnamefont {Moshe}},\ and\
  \bibinfo {author} {\bibfnamefont {I.}~\bibnamefont {Procaccia}},\ }\bibfield
  {title} {\bibinfo {title} {Anomalous elasticity and emergent dipole screening
  in three-dimensional amorphous solids},\ }\href
  {https://doi.org/10.1103/PhysRevE.107.055005} {\bibfield  {journal} {\bibinfo
   {journal} {Phys. Rev. E}\ }\textbf {\bibinfo {volume} {107}},\ \bibinfo
  {pages} {055005} (\bibinfo {year} {2023})}\BibitemShut {NoStop}%
\bibitem [{\citenamefont {Kumar}\ and\ \citenamefont {Procaccia}(2024)}]{24KP}%
  \BibitemOpen
  \bibfield  {author} {\bibinfo {author} {\bibfnamefont {A.}~\bibnamefont
  {Kumar}}\ and\ \bibinfo {author} {\bibfnamefont {I.}~\bibnamefont
  {Procaccia}},\ }\bibfield  {title} {\bibinfo {title} {Elasticity, plasticity
  and screening in amorphous solids: A short review},\ }\href
  {https://doi.org/10.1209/0295-5075/ad2087} {\bibfield  {journal} {\bibinfo
  {journal} {Europhysics Letters}\ }\textbf {\bibinfo {volume} {145}},\
  \bibinfo {pages} {26002} (\bibinfo {year} {2024})}\BibitemShut {NoStop}%
\bibitem [{\citenamefont {Cohen}\ \emph {et~al.}(2023)\citenamefont {Cohen},
  \citenamefont {Schiller}, \citenamefont {Wang}, \citenamefont {Dijksman},\
  and\ \citenamefont {Moshe}}]{23CSWDM}%
  \BibitemOpen
  \bibfield  {author} {\bibinfo {author} {\bibfnamefont {Y.}~\bibnamefont
  {Cohen}}, \bibinfo {author} {\bibfnamefont {A.}~\bibnamefont {Schiller}},
  \bibinfo {author} {\bibfnamefont {D.}~\bibnamefont {Wang}}, \bibinfo {author}
  {\bibfnamefont {J.~A.}\ \bibnamefont {Dijksman}},\ and\ \bibinfo {author}
  {\bibfnamefont {M.}~\bibnamefont {Moshe}},\ }\bibfield  {title} {\bibinfo
  {title} {Odd mechanical screening in disordered solids},\ }\href@noop {}
  {\bibfield  {journal} {\bibinfo  {journal} {arXiv preprint arXiv:2310.09942}\
  } (\bibinfo {year} {2023})}\BibitemShut {NoStop}%
\bibitem [{\citenamefont {Kaur}\ \emph {et~al.}(2025)\citenamefont {Kaur},
  \citenamefont {Procaccia},\ and\ \citenamefont {Samanta}}]{25KPS}%
  \BibitemOpen
  \bibfield  {author} {\bibinfo {author} {\bibfnamefont {P.}~\bibnamefont
  {Kaur}}, \bibinfo {author} {\bibfnamefont {I.}~\bibnamefont {Procaccia}},\
  and\ \bibinfo {author} {\bibfnamefont {T.}~\bibnamefont {Samanta}},\
  }\bibfield  {title} {\bibinfo {title} {Selection principle for the screening
  parameters in the mechanical response of amorphous solids},\ }\href
  {https://doi.org/10.1103/PhysRevE.111.015506} {\bibfield  {journal} {\bibinfo
   {journal} {Phys. Rev. E}\ }\textbf {\bibinfo {volume} {111}},\ \bibinfo
  {pages} {015506} (\bibinfo {year} {2025})}\BibitemShut {NoStop}%
\end{thebibliography}%
\end{document}